\documentclass[twocolumn,prl,showpacs]{revtex4}
\usepackage{graphicx,psfrag,dsfont,amsmath}
\usepackage{slashed,amssymb,color,pifont}
\def\bc{\begin{center}}
\def\ec{\end{center}}

\def\beq{\begin{equation}}
\def\eeq{\end{equation}}

\def\bk{{\bf k}}
\def\bp{{\bf p}}

\newcommand{\nn}{\nonumber}

\begin{document}

\title{Deformation of graphene sheet: Interaction of fermions with phonons}

\author{A. Sedrakyan$^{1,2}$, A. Sinner$^2$, and K. Ziegler$^2$}

\affiliation{$^1$Institute for Physics, Universit\"at Augsburg, Universit\"atsstr. 1,
D-86159, Augsburg, Germany\\
$^2$Yerevan Physics Institute, Br. Alikhanian 2, Yerevan 36, Armenia}

\begin{abstract}		
We construct  an effective low energy Hamiltonian which describes fermions dwelling on a deformed honeycomb lattice with dislocations 
and disclinations, and with an arbitrary hopping parameters of the corresponding tight binding model. 
It describes the interaction of fermions with a 2d gravity and has also a local $SU(2)$ gauge invariance of the group of rotations.
We reformulate the model as interaction of fermions with the deformation of the lattice, which forms a phonon field. 
We calculate the response of fermion currents to the external deformation or phonon field, which is a result of a $Z_2$ anomaly. 
This can be detected experimentally.     
\end{abstract}

\date{{\today}}

\maketitle

{\it Introduction:} 
The physics of electronic properties of strained~[\onlinecite{Kohmoto2006,montambaux2009,Pereira2009,Montambaux2009,EPL119,Phusit2018}] 
or lattice deformed graphene~[\onlinecite{Vozmediano2010,Vozmediano-2-2010,Vozmediano2012,Vozmediano2016,Volovik2014,Volovik-2015}] is an interesting problem, 
which reveals how concepts of 2d gravity can penetrate into the condensed matter area. 
Usually, one argues that deformations and strains give rise to the curvature of the surface of a 2d crystal, which is equivalent to the presence of gravity in a
two-dimensional world~[\onlinecite{Polyakov1981,Witten1983,Wiegmann2014}]. Moreover, it was argued by Vozmediano et al.~[\onlinecite{Vozmediano2010,Vozmediano2012,Vozmediano2016}], 
that besides the metric (or gravitational) field, a $U(1)$ gauge field shows up as well. In a recent paper~[\onlinecite{Radzihovsky-2018}] the dynamics of 
elastic deformations, dislocations and declinations of lattices was studied and an effective action for phonons was derived.    
The appearance of 2d gravity in the similar problems is not surprising. It is based on the paradigm that any reasonable definition of physical 
observables on random lattices should be covariant under the appointment of a coordinate system. In other words, the system should be 
reparametrization invariant, which leads to the emergence of 2d gravity. Moreover, any other degrees of freedoms based on distortions, 
declinations and irregularities should be governed by reparametrization invariance and
the fields describing them should have appropriate transformation properties. 

It is known from the differential geometry~[\onlinecite{Novikov1984}], that each random surface can be uniquely parametrized by a field of 
normal vectors ${\bf n}(\xi)$, where $\xi$ being elements of a two-dimensional coordinate system, and a three-component metric field $g_{\alpha\beta}$, 
which can be united to the so called conformal factor $\rho(\xi)$. The vector staying normal to the surface has two degrees of freedom, which together with $\rho(\xi)$ gives dual version of three degrees of freedom $\vec{X}(\xi)$ of the surface. 
Normal vector ${\bf n}(\xi)$ can be identified by the factor $SU(2)/U(1)$ of 3d rotations over $O(2)$ rotations around normal.
Therefore, one could expect, that the fermions living on surface should have the reparametrization (2d gravity) and 3d rotational symmetries. 
In Refs.[\onlinecite{Sedrakyan1985,Sedrakyan1987}] such theory of Dirac particles induced from the Clifford algebra in 3d was constructed.
The appearance of $SU(2)$ gauge symmetry essentially in the latter approach differs from the approach developed in 
Refs.~[\onlinecite{Vozmediano2010,Vozmediano-2-2010,Vozmediano2012,Vozmediano2016,Volovik2014,Volovik-2015}], 
where besides the gravity, only $U(1)$ gauge group is present.

In this paper we argue that the physics of fermions hopping with arbitrary parameters on a deformed honeycomb lattice can indeed 
be well described within the induced Dirac theory 
[\onlinecite{Sedrakyan1985,Sedrakyan1987}]. Considering deformations of the honeycomb lattice as an elastic field of phonons we 
reduce the problem to the interaction of fermions with phonons and define the corresponding Hamiltonian. We analyze the emerging 
$Z_2$ anomaly of this model and show how phonons may produce an anomalous current, which, in principle,
can be detected experimentally~[\onlinecite{Bostwick2020,Lemmens2020}].

\begin{figure}[t]
\centerline{\includegraphics[width=95mm,angle=0,clip]{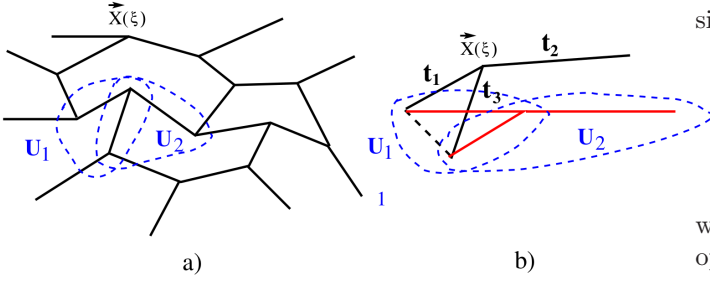}}
\caption{(Color online) 
(a) An example of random honeycomb lattice in 3d with two patches $U_1$ and $U_2$, which cover neighboring triangles. 
(b) Three-dimensional vertex $\vec{X}(\xi)$ (black) and	its projection on a flat space by rotation (red).}
\label{fig1}
\end{figure}

{\it Model for random deformations of the graphene sheet:} 
We depart from an arbitrary deformation of the honeycomb lattice in 2d space. For our consideration it is not important to have an exact honeycomb lattice. 
We consider a deformed surface which consists of sites with three attached links everywhere, while facets are not necessarily hexagons (there can 
be all possible $\bf n$-angles), see Fig.\ref{fig1} as an example. At each vertex we consider three independent hopping parameters 
$t^j(\xi),\; j=1,2,3 $ for the fermions with the Hamiltonian
\begin{eqnarray}
\label{H1}
H&=&\sum_{j,\xi}t^j(\vec{X})\big[\psi_A^+(\xi^\alpha+\mu_j^{\alpha}) \psi_B(\xi^\alpha)\nn \\
&+&\psi_B^+(\xi^\alpha) \psi_A(\xi^\alpha+\mu_j^{\alpha})\big]
\ ,
\end{eqnarray}
where notions A and B mark the natural partition of the honeycomb lattice into sublattices. 
Vector $\mu_j^{\alpha}$ connects neighboring sites on the parametric space and represents the difference of the coordinates
of neighboring sites in a patch. As for the manifolds we cover the whole lattice by a system of patches $U_a$, 
each of which envelops three neighboring sites. They may have an overlap region $ U_a \cap U_b $ covering neighboring links or a single site. 
An example of such coverings $U_1, U_2$ is presented in Fig.\ref{fig1}. Inside of each $U_a$ we have Cartesian coordinate systems,
which are connected by differentiable functions $\xi^{(a)}_\alpha=f^{ab}_\alpha(\xi^{(b)}_\beta)$.
This reparametrization transformation defines the gluing rules of the points in the overlap region. 
Because we are going to formulate a reparametrization invariant theory, it will have a well 
defined Hamiltonian, which depends on points, but not on the coordinate systems. This means also that we will have a 2d gravity theory. 

\begin{figure}[t]
\includegraphics[width=65mm,angle=0,clip]{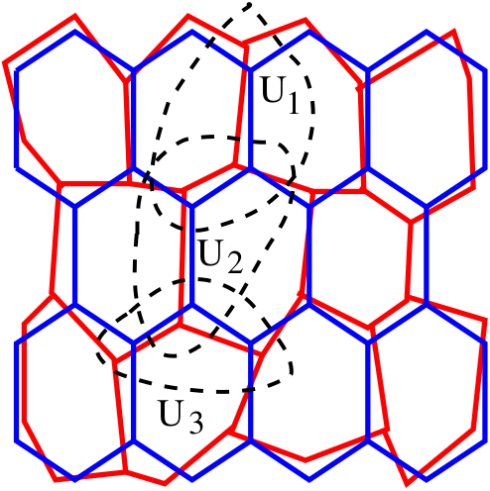}
\caption{(Color online) Flattened  random honeycomb lattice (red) and its reparametrization (deformation) to a regular one (red). 
Dotted black lines emphasize patches, which cover triangles.}
\label{fig:fig2}
\end{figure}

It is clear that by two local rotations in 3d along the hopping links we can make triangles in each patch parallel to the
$(x,y)$-plain.  Each of such links contains a pair of fermions $\Psi(\xi)=(\psi_A(\xi),\psi_B(\xi))$ in the corresponding $U_a$,
which form a spinor representation $\Omega_a(\xi)$ of rotation group $SU(2)$. Rotations on different patches are different, but 
on the overlap region they are connected by rotations  $\Omega_{ab}=\Omega^{-1}_a(\xi^{(a)})\Omega_b(\xi^{(b)})$,
which gives rules of gluing of tangent vectors on different patches. In parallel to a reparametrization
symmetry, our Hamiltonian should also have a local gauge $SU(2)$ symmetry.
In Fig.\ref{fig:fig2} we show the flat projection of the random lattice surface in 3d (marked red), which 
can be reparameterized as a regular honeycomb lattice (marked blue). Black dotted lines emphasize the open disc 
patches of Cartesian coordinate systems. 

After a rotation our Hamiltonian (\ref{H1}) in a 2d basis space becomes
\begin{eqnarray}
\label{H2}
H=\frac{1}{2}\sum_{j,\xi}t^j\Psi^+(\xi)\Omega^+(\xi)\sigma_1\Big[e^{ \sigma_3\mu_j^{\alpha}\overrightarrow{\partial}_\alpha} \nn\\
+ e^{- \overleftarrow{\partial}_\alpha\mu_j^{\alpha}\sigma_3} \Big]\Omega(\xi)\Psi(\xi),
\end{eqnarray}
where left/right arrows above the partial derivative operators point into the direction of their action and $\sigma_{1,3}$ are Pauli matrices. 
Since there is a deformed honeycomb structure on a plain we expect the existence of a local, patch dependent momentum $K_\alpha$, such that
\begin{equation}
\label{zero-point}
\sum_{j} t^j e^{i \mu_j^\alpha K_\alpha} =0.
\end{equation}
This condition defines two real equations for the momenta $K_\alpha$ and lattice vectors $\mu_j^\alpha$ in each patch $U_a$. 
Then, in order to consider the low lying excitations around these points, 
which are of our primary interest, we shift derivatives in the exponents in (\ref{H2}) by $\varphi_j= \mu_j^\alpha K_\alpha $ and replace  
$\mu_j^{\alpha} \overrightarrow{\partial}_\alpha \rightarrow i \varphi_j+\mu_j^{\alpha} \overrightarrow{\partial}_\alpha$ and $-\mu_j^{\alpha} \overleftarrow{\partial}_\alpha \rightarrow i \varphi_j-\mu_j^{\alpha} \overleftarrow{\partial}_\alpha$. 
By doing this and taking into account that vectors $\mu_j^\alpha$ are proportional to the minimal length scale of the lattice $\varepsilon$, 
we can expand the translation operators $ e^{-\overleftarrow{\partial}_\alpha\mu_j^{\alpha}}$ and  
$ e^{\overrightarrow{\partial}_\alpha\mu_j^{\alpha}}$ and keep only linear terms. 
In order to expand the exponent one should first decouple in the exponential term $\varphi_j$ from the
derivatives by using the Campbell-Hausdorff formula~[\onlinecite{Novikov1984}]. 
Then a commutator term will appear. However, the commutator terms from the two exponents cancel each other. 
Eventually, the Hamiltonian of the low energy states becomes
\begin{eqnarray}
\label{H3}
\!H\!&=&\!\frac{i}{2}\sum_{j,\xi}\Psi^+(\xi)\Omega^+(\xi) \Big(t^j\sigma_2 \cos[\varphi_j]-t^j\sigma_1 \sin[\varphi_j]\Big)
\nn\\
\!&\cdot&\! \Big[\mu_j^{\alpha}\overrightarrow{\partial}_\alpha-\overleftarrow{\partial}_\alpha\mu_j^{\alpha}
\Big] \Omega(\xi)\Psi(\xi),
\end{eqnarray}
where Eq. (\ref{zero-point}) was used. Denoting now the coefficients of the Pauli matrices in (\ref{H3}) as
\begin{eqnarray}
\label{tetrad}
\varepsilon e^{\alpha 2}&=&\sum_{j} t^j(\vec{X})\mu^{\alpha}_j \cos[\varphi_j]\nn,\\
\varepsilon e^{\alpha 1}&=&-\sum_{j} t^j(\vec{X})\mu^{\alpha}_j \sin[\varphi_j],
\end{eqnarray} 
we can regard  $e^{\alpha a}, a=1,2$ as tetrads in 2d gravity and the Hamiltonian 
of fermions on the fluctuating surface becomes
\begin{equation}
\label{H4}
\!\!H=\frac{i}{2}\sum_{\xi}\varepsilon e \Psi^+(\xi)\Omega^+(\xi) e^{\alpha a}\sigma_a \big(\overrightarrow{\partial}_\alpha-\overleftarrow{\partial}_\alpha\big)\Omega(\xi)\Psi(\xi).
\end{equation}
Here $\varepsilon$ is the minimal length scale of the lattice. By using an ambiguity
of the coordinate vectors $\mu^\alpha_j$ one can associate tetrads $e^{\alpha a}$
with the induced metric of the surface $g_{\alpha\beta}=\partial_\alpha\vec{X}\partial_\beta\vec{X}$.
Namely, we can fix the coordinate vectors $\mu^\alpha_j$ in such a way that
\begin{equation}
\label{metric}
\partial_\alpha\vec{X}\partial_\beta\vec{X}=\sum_{a=1,2}e^a_\alpha e^a_\beta,
\end{equation}
where $e^a_\alpha=[e^{\alpha a}]^{-1}$, defined in Eq.(\ref{tetrad}). From the formulas (\ref{tetrad})
and (\ref{metric}) it is clear, that in an appropriate parametrization of the random lattice sites the length of the vector's tangent 
of the surface is proportional
to the corresponding hopping parameter $t_j$, e.g. $\mu_j^\alpha |\partial_\alpha \vec{X}|\simeq \varepsilon t_j$. 
In total, there are two equations (\ref{zero-point}) and three equations (\ref{metric}) to be solved,
from which 6+2=8 free parameters of $\mu_j^\alpha, j=1,2,3$ and $K_{\alpha}, \alpha=1,2$ must be determined. 
Hence we will get 3 parameter solutions, which reflects d=3 free parameters of the surface coordinates $\vec{X}(\xi)$, 
or equivalently, 3 parameters of the induced metric $g_{\alpha\beta}$.

Hamiltonian (\ref{H4}) coincides with the Hamiltonian of the Dirac theory on 2d random surfaces induced from 3d flat Dirac theory 
with an Euclidean metric defined in Refs.~[\onlinecite{Sedrakyan1985,Sedrakyan1987}]. It was shown that by defining the
induced gamma matrices as 
$\hat{\gamma}_\alpha = \partial_\alpha \vec{X} \vec{\gamma}$ ($\vec{\gamma}$ are 3d Dirac $\gamma$-matrices) and  
a 3d rotation one arrives at the simpler Hamiltonian
\begin{equation}
\label{H5}
H=\frac{i}{2}\int d\vec{\xi} \sqrt{g } \Psi^+(\xi) \hat{\gamma}^\alpha \big(\overrightarrow{\partial}_\alpha-\overleftarrow{\partial}_\alpha\big)\Psi(\xi)
\ ,
\end{equation}
where $g=\det[g_{\alpha\beta}]$. This expression shows that besides 2d gravity we have also local 3d rotations, which induce a
non-Abelian $SU(2)$ gauge field. Transforming left differential in (\ref{H5}) to the right one, we obtain
\begin{equation}
\label{H6}
H=i \int d\vec{\xi} \sqrt{g} \Psi^+(\xi) \big(\hat{\gamma}^\alpha \partial_\alpha +\frac{1}{2}\nabla_\alpha \hat{\gamma}^\alpha \big)\Psi(\xi)
\ ,
\end{equation}
where $\nabla_\alpha $ is a covariant derivative defined by Christoffel symbols~~[\onlinecite{Novikov1984}]. The term $ \nabla_\alpha \hat{\gamma}^\alpha = \sqrt{g} h_{\alpha}^\alpha \hat{n }$ 
is connected with the second quadratic form
$h_{\alpha \beta}=\vec{n} \nabla_\alpha \partial_\beta \vec{X}$, where $\vec{n}=\sqrt{g}\partial_1\vec{X} \times \partial_2 \vec{X} $ is the
vector normal to the surface at $\xi$. In Ref.~[\onlinecite{RandLatt2011}], the Hamiltonian (\ref{H6}) was used to calculate
optical conductivity of fermions on a random surface.

Hamiltonian (\ref{H6}) can be written as a 2d gravity theory of fermions, 
which interacts with a non-Abelian $SU(2)$ gauge field of rotations, cf. [\onlinecite{Sedrakyan1985,Sedrakyan1987}]. Introducing rotations
\begin{equation}
\label{Omega}
\Omega(\xi)=\frac{1+\hat{m}_{a}(\xi) \hat{m}_{a}(0)+\hat{n}(\xi)\hat{n}(0)}{2 \sqrt{1+{\bf m}_{a}(\xi)\cdot{\bf m}_{a}(0)+{\bf n}(\xi)\cdot{\bf n}(0)}},
\end{equation}
where ${\bf m}_{a}(\xi)\;(a=1,2)$ are the two tangential and ${\bf n}(\xi)$ the normal vectors at ${\bf \xi}$,
with  $\hat{m}_{a}(\xi)={\bf m}_{a}{\bf \sigma}$ and $\hat{n}(\xi)={\vec n}(\xi){\vec \sigma}$ ,
one can write
\begin{eqnarray}
\label{H7}
H&=&i \int d\vec{\xi} \sqrt{g} \Psi^+(\xi) \hat{\gamma}^\alpha \Big( \partial_\alpha \nn\\
&+&\Omega^+(\xi) \partial_\alpha \Omega(\xi) +\frac{1}{2}\hat{n} \omega_\alpha \Big)\Psi(\xi).
\end{eqnarray}
Here $\omega_\alpha=\frac{1}{2}e^\beta_a \nabla_\alpha e_{\beta a}$ is the spinor connection.

\begin{figure}[t]
\center
\includegraphics[width=45mm,angle=0,clip]{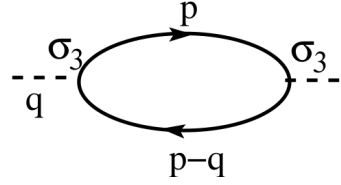}
\caption{(Color online) One loop Feynman diagram for $\langle j^{}_3j^{}_3\rangle$ correlator.}
\label{fig3}
\end{figure}

{\it Phonon -  fermion interaction:} Our goal is to understand how phonons interact with
fermions on a graphene sheet defined by the Hamiltonian~(\ref{H6}). The phononic field is the field
of elastic deformations of the graphene sheet $\vec{X}(\xi)$~[\onlinecite{Vozmediano2016,Basko2008}]. 
On a flat regular honeycomb lattice background we write
\begin{equation}
\label{phonon field}
\vec{X}(\xi)= \xi^a \vec{a}_{a} + \vec{u}(\xi)
\end{equation}
where $\vec{a}_a, (a=1,2)$ are two basic vectors on a flat plane and $\vec{u}(\xi)$ is the phonon field.
The differential operator ${\cal D}= \sqrt{g}\hat{\gamma}^\alpha \partial_\alpha +\frac{1}{2}\nabla_\alpha \hat{\gamma}^\alpha $, which appears
between fermionic fields in the Hamiltonian in lowest order of the phonon field, acquires the form
\begin{eqnarray}
\label{H8}
H&=&\int d\vec{\xi} \sqrt{g} \Psi^+(\xi)\;{\cal D}\;\Psi(\xi)\nn\\
{\cal D}&=&i\sigma_a \partial_a + i T_{j a} \sigma_j \partial_a +\sigma_a A_a
+\sigma_3 M.
\end{eqnarray}
Except for the middle term with the stress energy tensor
\begin{equation}
\label{T2}
T_{ja}=-\partial_a u^j,
\end{equation}
the covariant derivatives of Eq.~(\ref{H8}) coincide with those considered in the model of electron-phonon interaction in Refs.~[\onlinecite{Phonon2016}] 
and [\onlinecite{Phonon2019}].  Here and below the repeated indices denote summations over $a,b=1,2$ and $i,j =1,2,3$, respectively, while
\begin{eqnarray}
\label{A1}
M\!&=&\! ~\frac{1}{2} \Big[\partial^2_a u^3+ \partial_au^a \partial^2_a u^3- \partial^2_au^b u^3_b\Big]\\
\label{A2}
A_a\!&=&\!-\frac{1}{2} \Big[\partial_a u^3 \partial^2_b u^3+ \partial^2_bu^\sigma(\partial_a u^\sigma+
\partial_\sigma u^a) \Big].\label{A2-1}
\end{eqnarray}
Formally $M$ can be considered as a mass term, while $A_a,\;a=1,2$ are components of a $U(1)$ gauge field that emerged due to the deformations of the honeycomb lattice. 
One recognizes that lowest order in the phonon field $u$ contributes only
into the mass term, while the emerging $U(1)$-gauge field appears in quadratic over $u$ order.
Therefore, to leading order of ${\cal O}(u^1)$ the fermionic fluctuations create an effective action $S^{}_N(M)$ based on the correlator between 
the 'axial currents' $\langle j_3 j_3 \rangle$. Adopting the dimensional regularization scheme one obtains to one loop order, cf. Fig.~\ref{fig3}
~[\onlinecite{Redlich1984,Son2007, apresyan-2017}] 
\begin{equation}
\label{W} 
S_N(M)=\frac{1}{8}\int \frac{d^3k}{(2\pi)^3} \sqrt{k_0^2+ {\bf k}^2} ~ M_{k}M_{-k}
\end{equation}
written in momentum space. In deriving this expression one has to keep in mind that $\sigma_3$ does not commute with the fermionic propagator, cf. supplement materials.

In the next order in $u$'s, i.e. to the order ${\cal O}(u^2)$, there appear anomalous current-current correlators $\langle j_a j_b \rangle$ corresponding 
to the remaining two space-like components of the gauge field $A^{}_{a=1,2}$. According to the seminal works of  Redlich~[\onlinecite{redlich84,Redlich1984}],
Semenoff~[\onlinecite{Semenoff-1984}] and Jakiw~[\onlinecite{Jackiw}], the
\begin{equation}
\label{W2} 
S^{}_A(A) = -i{\rm sgn}(m)\epsilon_{ab} \int d\tau d^2x ~ A^{}_a  \partial_\tau A^{}_{b},
\end{equation}
where ${\rm sgn}(m)$ refers to an infinitesimally small  bare mass parameter $m$, which was introduced to regularize the infrared
divergence and sent to zero after the integration. Variation of the action Eq.~(\ref{W2}) creates an anomalous current
\begin{equation}
\label{J}
j_a= -i{\rm sgn}(m)\epsilon_{ab} \partial_\tau A_{b}.
\end{equation}
The sign ($Z_2$) ambiguity of the mass reflects the fact that the mass parameter must not necessarily be positive.   
As it is always the case with anomalies in perturbative approaches, the anomalous current Eq.~(\ref{J}) appears because 
the regulator violates the chiral symmetry of the model Eq.~(\ref{H7}) explicitly. 
Due to the finite band width of our lattice model, there is no need for an ultraviolet regularization.

In this case the chiral symmetry is preserved and the anomalous currents cancel each other
due to fermion species doubling~\cite{susskind77}. However, if the dynamics of phonons is included into the model, 
the breaking of the chiral symmetry can 
occur spontaneously, provided the phonon-phonon interaction strength exceeds a certain critical value. This mechanism was recently 
investigated by two of us in Refs.~[\onlinecite{Phonon2016,Phonon2019,Phonon2020}]. 

{\it Conclusions:} 
In the present paper we  construct a low energy theory of fermions interacting with deformations of the honeycomb lattice.
In contrast to similar studies reported recently in 
Refs.~[\onlinecite{Vozmediano2010,Vozmediano-2-2010,Vozmediano2012,Vozmediano2016,Volovik2014,Volovik-2015}],
where fermions are bound to the flat but distorted sheets, we investigate the case when the effective gauge fields are induced by 
embedding of a two-dimensional surface into a three-dimensional Euclidean space~[\onlinecite{Sedrakyan1985,Sedrakyan1987}]. 
In addition to the $U(1)$ gauge fields and interaction with 2d gravity of former approaches, our effective theory reveals a
non-abelian $SU(2)$ gauge field. 
We reduce the 2d gravity (metric) field to deformations of the 3d lattice, which is forming three
dimensional phononic fields. The calculation of a $Z_2$ anomaly links the current of fermions with
phononic field strength which, in principle, can be detected experimentally.    
It remains for the future to extend the formalism presented here
to the curved spaces. The ultimate goal may be to establish an effective low-energy field theory of phonons in 
the spirit of effective Liouville actions~[\onlinecite{Polyakov1981,Polyakov1983}] accompanied 
by induced topological (Chern-Simons or Hopf) terms. To some extend though, a number of intermediate ideas in terms of mathematical modeling 
and its effect on transport were successfully realized in [\onlinecite{RandLatt2011}]. 

{\it Acknowledgments}.
A. Sedrakyan expresses his gratitude to the University Augsburg for hospitality during his stay as visiting professor, 
where this work was initiated. His work was further supported through the ARC grants 18RF-039 and 18T-1C153.
A. Sinner and K. Ziegler were supported by the grants of the Julian Schwinger Foundation for Physics Research.

\pagebreak
\widetext
\begin{center}
	\textbf{\large Supplemental Materials: Deformation of graphene sheet: Interaction of fermions with phonons}
\end{center}
\setcounter{equation}{0}
\setcounter{figure}{0}
\setcounter{table}{0}
\setcounter{page}{1}
\makeatletter
\renewcommand{\theequation}{S\arabic{equation}}
\renewcommand{\thefigure}{S\arabic{figure}}
\renewcommand{\bibnumfmt}[1]{[S#1]}
\renewcommand{\citenumfont}[1]{S#1}

\section*{ Derivation of Eq.~(16)}

Writing the diagram depicted in Fig.3 as an algebraic expression yields:
\begin{eqnarray}
\nn
\langle j_3 j_3 \rangle &\sim& {\rm tr}\Big[M\sigma^{}_3[\partial^{}_\tau\sigma^{}_0 + i\partial^{}_\nu\sigma^{}_{\nu=1,2}]^{-1}\Big]^2  \\
&=&
\int\frac{d^3P}{(2\pi)^3}M_{P} M_{-P} 
\underbrace{{\rm Tr}~\int\frac{d^3K}{(2\pi)^3}\frac{\sigma^{}_3[ik^{}_0\sigma^{}_0+k^{}_\nu\sigma^{}_\nu]\sigma^{}_3[i(k^{}_0+p^{}_0)\sigma^{}_0+(k^{}_\mu+p^{}_\mu)\sigma^{}_\mu]}{K^2(K+P)^2}}_{=\Pi^{-1}_{33}(P)}
\end{eqnarray}
where $K^2=k^2_0+\bk^2$. With the Feynman trick 
\begin{equation}
\frac{1}{AB}=\intop_0^1~\frac{dx}{[(1-x)A+xB]^2},
\end{equation}
follows
\begin{equation}
\Pi^{-1}_{33} (P) = {\rm Tr}\int_0^1 dx~\int\frac{d^3K}{(2\pi)^3}\frac{[ik^{}_0\sigma^{}_0-k^{}_\nu\sigma^{}_\nu][i(k^{}_0+p^{}_0)\sigma^{}_0+(k^{}_\mu+p^{}_\mu)\sigma^{}_\mu]}{[(1-x)K^2+x(K+P)^2]^2},
\end{equation}
where the sign change in the first term is because $\sigma^{}_\nu$ anticommutes with $\sigma^{}_3$. Shifting $k^{}_{i=0,1,2}\to k^{}_i - x p^{}_i$, performing the trace and keeping only rotationally invariant terms yields
\begin{equation}
\Pi^{-1}_{33} (P) = 2\int_0^1 dx~\int\frac{d^3K}{(2\pi)^3}\frac{x(1-x)P^2 - K^2}{[K^2+x(1-x)P^2]^2}.
\end{equation}
Using formulas of dimensional regularization
\begin{eqnarray}
\label{eq:DimReg}
\intop\frac{d^dK}{(2\pi)^d} \frac{1}{(K^2+\Delta)^n} &=& \frac{1}{(4\pi)^{d/2}}
\frac{\Gamma(n-\frac{d}{2})}{\Gamma(n)\Delta^{n-{d/2}}}, \\
\nonumber
\intop\frac{d^dK}{(2\pi)^d} \frac{K^2}{(K^2+\Delta)^n} &=& \frac{1}{(4\pi)^{d/2}}\frac{d}{2} \frac{\Gamma(n-\frac{d}{2}-1)}{\Gamma(n)\Delta^{n-{d/2}-1}},
\end{eqnarray}
yields 
\begin{equation}
\Pi^{-1}_{33} (P) = \frac{1}{\pi}|P|\int_0^1 dx \sqrt{x(1-x)} = \frac{1}{8}\sqrt{p^2_0+\bp^2}, 
\end{equation}
where the integration over $x$ is carried out with the substitution $x=\sin^2\alpha$, giving 
\begin{equation}
\int_0^1 dx~\sqrt{x(1-x)} = \int^{\pi/2}_{0}d\alpha~\sin^2\alpha\cos^2\alpha=\frac{\pi}{8}. 
\end{equation}

\end{document}